# Surface Oxidation of the Topological Insulator $Bi_2Se_3$

*Avery J. Green*, Sonal Dey, Yong Q. An, Brendan O'Brien, Samuel J. O'Mullane, Bradley Thiel, and Alain C. Diebold*
*Colleges of Nanoscale Science and Engineering, SUNY Polytechnic Institute,*
*257 Fuller Road, Albany, NY 12203*
**e-mail address: agreen@sunypoly.edu*

A comprehensive picture of the aging and oxidation of the (0001) surface of $Bi_2Se_3$ is critical to understanding the physical origin of changes in its topologically protected surface states. We find that surface aging in ambient conditions occurs in two major steps. Within two hours of exfoliation, a series of ~ 3.6 Å high islands are observed by atomic force microscopy over approximate 10% of the surface. Subsequently, patch growth stops, and oxidation begins after the two hours and continues until one quintuple layer has been oxidized. X-ray photoelectron spectroscopy shows no sign of oxidation before ~ 120 minutes of exposure to air, and the oxygen 1s peak is clearly present after ~ 190 minutes of ambient exposure. Variable angle spectroscopic ellipsometry also indicates that the oxidation of a full quintuple layer occurs on the time scale of days. These results are in good agreement with the time dependent changes previously observed in the surface crystal structure by second harmonic generation.[1,2] In addition to providing the ability to non-destructively measure oxide on the surface of $Bi_2Se_3$ crystals, ellipsometry can be used to identify the thickness of $Bi_2Se_3$ flakes. These results are consistent with non-linear optical methods including rotational anisotropy second harmonic generation, which follow the time dependence of the changes to the top quintuple layer.[2]

## I. INTRODUCTION

Since the experimental observation of the quantum Hall effect by von Klitzing *et al.*[3] and its theoretical explanation by Thouless *et al.*,[4] nontrivial electronic topologies in insulating crystals have been studied as novel material phases. Following the development of lab-scale 2D sample production methods in the search for graphene[5] and the discovery of time-reversal invariant $Z_2$ topological insulators (TIs),[6] there has been a considerable effort to experimentally realize the unique carrier properties of TI crystals in a variety of structures and devices. Aside from the challenge of producing measurable samples, this effort has been met with the challenge of distinguishing between conductance from bulk states, topological surface, and more recently, 2D electron gas (2DEG) states.[7,8] Thus, the study of the long term time dependence of surface aging and oxidation is a critical part of understanding the impact of significant changes in the surface on the topologically protected states. For example, oxidation can change the quintuple layer (QL) structure by partially oxidizing the top layer. A number of changes in the surface structure are possible including segregation-induced changes in the atomic layer structure (aging), oxidation, contamination, doping near the surface due to aging or oxidation, structural defects, and other effects both independent of, and correlated with these. Although surface aging clearly impacts the properties of topological insulators, a comprehensive experimentally determined understanding of surface aging after exfoliation of a clean surface is missing from the literature. In addition, the literature has inconsistencies in the reported time dependence of key aspects such as surface oxidation and Bi diffusion. This paper reports a comprehensive picture of the time dependence of surface aging of a key TI, $Bi_2Se_3$, based on a variety of systematically applied characterization methods.

The semiconducting alloy $Bi_2Se_3$ serves as a prototypical material for demonstrating the spin-momentum locking because of its relatively large bulk band gap (0.3 eV), and its simple surface band structure (single Dirac cone).[9] The surface states and the related surface structure of $Bi_2Se_3$ have been extensively studied by various analytical techniques, such as angle-resolved photoemission spectroscopy measurement of the valence band structure (ARPES),[7,10–12] x-ray photoelectron spectroscopy (XPS) including angle resolved XPS (ARXPS) of the surface elemental composition, chemical state and surface oxidation,[13–15] low energy electron diffraction (LEED) and surface x-ray diffraction (SXRD),[16] ion scattering,[17] and scanning tunneling microscopy.[18] Significant discrepancies in surface composition have been reported for vacuum cleaved (0001) $Bi_2Se_3$ surfaces. Note that the (0001) plane is in reference to the hexagonal unit cell, for which we use Miller-Bravais 4-axis notation, and that this system is substitutable with the geometrically distinct rhombohedral unit cell system, for which the same surface plane is (111). For example, a bismuth bilayer was observed at room temperature by using ion scattering along with thermally activated segregation of bismuth to the surface at 80K while LEED measurements show no signs of the Bi bilayer.[17] This discrepancy is likely due to differences in the $Bi_2Se_3$ crystal. Bi bilayers in the bulk between the QLs were observed by aberration-corrected scanning transmission electron microscopy (STEM) for some $Bi_2Te_3$ crystals,[19] and it is likely that some crystals of $Bi_2Se_3$ will also have bismuth bilayers in between QLs. High resolution aberration-corrected microscopy studies confirm the variability of bulk tetradymite chalcogenide crystals, and here we confirmed the quality of the $Bi_2Se_3$ sample used in this study using aberration corrected STEM.

Although a number of publications describe the results of characterization of $Bi_2Se_3$ surfaces that were cleaved and maintained in ultra-high vacuum, many other electrical and optical experiments were performed in air to facilitate sample manipulation and signal acquisition. Some of these studies also characterized the surface after exposure to air. Humidity present in the surrounding air and the duration of exposure of the fresh surface to the surrounding air are both critical to preservation and detection of the surface states. Several studies describe the effect of short-term exposure to air or dry nitrogen.[15,20] Selenium vacancies can form after exfoliation, which dopes the sample surface. For example, ARPES characterization found that the topologically protected states remain robust, but a 2D electron gas is formed at the surface due to doping-induced band bending.[10] This data shows that even for short exposures to nitrogen or air, the position of the Dirac cone is shifted relative to the Fermi level due to electron doping of the surface states. Longer term XPS characterization of the time dependent oxidation of $Bi_2Se_3$ has shown that moisture impacts oxidation process.[21] One study reports the formation of a thin surface oxide of ~ 0.39 nm thickness on the surface of Sn-doped $Bi_2Se_3$ nanoribbon after minimum air exposure.[22] After two days of exposure, this study estimates the oxide thickness to be ~ 2 nm for the bulk crystal. The oxide was identified using the Bi 4f spectra. The line shape of many heavy metal f core level peaks is complicated by shake-up events during electron emission.[23,24] Since the line shape of Bi 5d peaks are not altered by shake-up, Bi 5d peaks are preferable for observation of the initial stages of oxidation in $Bi_2Se_3$. The Se 3p peaks occur at a slightly higher binding energy than the Bi 4f peaks, resulting in a change of the apparent peak shape of the Bi 4f 5/2, 7/2 doublet. Despite this altered peak shape, the effect of shake-up events on oxide free Bi 4f state peak shape is readily observed in $Bi_2Te_3$ in the 0 hour data of Figs. 3(a) and 3(b) of reference[25] and also in our measurements of the Bi 4f region (see appendix). Another ARPES study reported that $H_2O$ exposure results in band bending due to n type doping as a result of positive Se vacancies which form after $Bi_2Se_3$ releasing Se.[20] This same study reports that vacuum cleaved $Bi_2Se_3$ remains undoped after three days in ultra-high vacuum. Another source used ARPES to find doping dynamics that caused the Dirac point to shift toward or away from the Fermi energy, depending on different cleaving-induced surface characters.[26]

A smooth surface morphology is important to the properties of heterostructures of $Bi_2Se_3$ with other materials. This study reports the time dependent formation of surface islands using atomic force microscopy. Islands are first observed after 33 minutes after surface exfoliation, and only minimal island formation is seen after ~100 minutes. The ~0.36 nm high islands cover less than 10% of the surface area. The authors are not aware any previous reports of this phenomena. The islands do not seem to alter the results of surface sensitive methods that sampling areas larger that the islands. Variable angle spectroscopic ellipsometry (VASE) and XPS are sensitive to time dependent changes due to surface morphology and oxide growth of the undoped (0001) $Bi_2Se_3$ surface. The XPS and AFM results provide justification for the optical models of the surface structure presented here.

Non-linear optical methods provide evidence of the continued existence of topologically protected states for (0001) $Bi_2Se_3$ surfaces exposed to air for extended periods.[27–29] Rotational anisotropy second harmonic (RA-SH), circular dichroism SH, and pump-probe methods have been used to characterize the surface states of As doped (0001) $Bi_2Se_3$ where the As doping reduces the bulk carrier concentration.[27–29] Polarization dependent rotational anisotropy second harmonic (RA-SH) generation probes the surface of freshly cleaved (0001) $Bi_2Se_3$ as well as the accumulation region after oxidation induced band banding. Second harmonic circular dichroism (CD) was used to verify the time reversal symmetry of the topologically protected surface states of (0001) $Bi_2Se_3$.[28] No differences were observed between the surfaces of as-cleaved and samples aged for 200 minutes in that study. The dependence of RA-SH on the number of QLs of $MgF_2$ capped (0001) $Bi_2Se_3$ surfaces has also been studied.[30] Reflectance pump-probe characterization has been used to characterize both carrier and phonon dynamics of undoped (0001) $Bi_2Se_3$ surfaces.[31]

However, the time dependent changes in the surface structure and surface states have only recently been studied by non-linear optical methods.[2] The short and long term time dependence of the surface structure and the carrier dynamics of dry air and dry nitrogen exposed, undoped samples of $Bi_2Se_3$ show that RA-SH is sensitive to the effects of oxidation inside the top QL.[2] In our study, circularly polarized pump and linearly polarized pump-SH probe detected spin polarization of photoexcited surface states on freshly cleaved $Bi_2Se_3$ surfaces, but not on oxidized surfaces. Thus, time dependence of the sample surface as well as the amount of humidity and doping are critical to the impact of ambient exposure on surface states. Other materials such as silicon are known to show dopant dependence to the oxidation rate. The results of our recent RA-SH and pump-probe measurements are further interpreted in terms of the time dependence of the surface structure observed by angle resolved XPS and VASE.

Here, we report a comprehensive picture of the surface aging of the cleaved (exfoliated) surface of $Bi_2Se_3$. This paper is structured as follows. First, the details regarding the $Bi_2Se_3$ crystal used in this study and the experimental methods are presented in section II. Next, the QL structure and lack of Bi bilayers between QLs is verified using aberration corrected STEM and X-ray diffraction in section III.A. The observation of island formation using AFM is presented in section III.B. Section III.C describes time dependent oxidation of (0001) $Bi_2Se_3$ including the first signs of Bi and Se oxidation after ~190 minutes using XPS and ARXPS. In section III.D, dielectric functions of the bulk crystal and surface oxide are measured with VASE, and oxide growth is non-destructively tracked. Section III.E shows the non-linear optical properties of $Bi_2Se_3$ and describes a change in surface crystal symmetry with second harmonic generation (SHG). A conclusion (section IV) follows to summarize the findings.

## II. EXPERIMENTAL METHODS

Ex-situ cleaved $Bi_2Se_3$ crystals from Sigma Aldrich were used for all of the measurements. Variable angle spectroscopic ellipsometry data was taken over a wavelength range from 245 nm to 1700 nm using a J.A. Woollam Co. dual rotating compensator ellipsometer tool (RC2). An elliptical measurement area of minor axis length of 250 μm is measured at all angles by the focusing optics used in this experiment.

High resolution X-ray diffraction (HR-XRD) data was collected using a Jordan Valley – Bede Metrix-L diffractometer using the Cu $K_\alpha$ line. The size of the x-ray beam used was 1x1 mm. The Omega-2Theta scan that produced (0 0 0 3n) information was performed in air, took data for 2 seconds per point, and had an angular step size of 0.005°.

Surface topography data was collected using a Bruker Dimension 3100 SPM in tapping atomic force microscopy mode. The AFM's atmospheric environment and short (<20 minute) capture time per each frame make the dynamic topography of the surface readily visible.

XPS and ARXPS data was collected using a Thermo Scientific Theta Probe™ system using monochromatic Al $K_\alpha$ (1486.6 eV) x-rays of 400 μm spot-size. With the exception of the t = 0 minutes sample, all other samples for the XPS study were exfoliated in ambient air. The t=0 sample was exfoliated inside a portable glove-box attached on top of the load lock under $N_2$ rich environment. All samples were loaded simultaneously into the ultra-high vacuum (UHV) chamber and the measurements were made at a base pressure of ≈ $8.0\times10^{-9}$ Torr. The measurements were taken with an energy step of 0.05 eV and at pass energy settings of 100 eV and 50 eV for Bi 5d and Se 3d peak regions respectively. The energy resolution is estimated to be 1.0 eV from the full width at half maximum of the Au $4f_{7/2}$ core level. In the absence of a well defined C 1s peak after exfoliation, the energy scale of the XPS spectra was calibrated with respect to the Se $3d_{5/2}$ core level (53.7 eV) according to the value in the NIST database.[32] We estimate the uncertainty with all peak binding energies to be ± 0.05 eV from peak fitting. ARXPS data were measured by collecting photoelectrons in 16 channels over an angular range of 24.875° to 81.125° using a spherical sector analyzer fitted with an electrostatic lens having 60° angular acceptance and an angular dispersive multichannel detector.

Cross-sectional aberration-corrected scanning transmission electron microscopy (STEM) was taken with an FEI Titan microscope operating at 300 keV.

Second Harmonic generation was excited by a pulsed laser at a wavelength of 740 nm, a pulse duration of ~120 fs, and a repetition rate of 82 MHz. The incident beam was chopped at the frequency of 3.3 kHz to reduce laser heating so as to minimize the sample surface damage. The incident beam was at a 45° incident angle and focused to a spot about 30 μm in diameter on the sample, resulting in a typical fluence of about 35 μJ/cm$^2$. A dry air jet was employed to further alleviate the heating issues inherent to the nonlinear optical experiments using an intense laser beam.

## III. RESULTS AND DISCUSSION

### 1. Verification of Defect Free QL Structure

In order to establish the quality of the $Bi_2Se_3$ crystal, the QL structure was verified using aberration-corrected STEM and HR-XRD. The primitive cell of a $Bi_2Se_3$ crystal is rhombohedral, and of space group R$\bar{3}$m (166). This primitive cell is geometrically distinct from the commonly used hexagonal unit cell of $Bi_2Se_3$, which contains the volume of three primitive cells. In the hexagonal unit cell, atoms are arranged in two-dimensional planes normal to the [0001] direction, such that each plane consists of one element. These planes have a rhombohedral vertical stacking arrangement. Every group of five planes is one QL, and all QLs have elemental planes with a $Se_{(2)}^{(A)}$-$Bi^{(B)}$-$Se_{(1)}^{(C)}$-$Bi^{(A)}$-$Se_{(2)}^{(B)}$ vertical arrangement (Fig. 1). Each QL itself has a rhombohedral horizontal translation, thus placing the other two mid-QL $Se_{(1)}$ atoms in the unit cell at A and B sites, and is separated by a relatively weak van der Waals (vdW) gap. As such, the material can be cleaved easily (using a scotch tape, for example) along the [0001] direction to produce a fresh (0001) surface of the $Bi_2Se_3$ crystal.

This complex bulk crystal structure is known to be susceptible to various crystal defects including the presence of Bi bilayers between QLs,[19] point defects originating from Se vacancies[33] and twinning.[34] In order to verify the atomic structure of the crystal used in this study, cross-sectional and top-down images of the crystal structure were obtained using HAADF STEM as shown in Fig. 2. This sample has a well-ordered QL structure, and Bi bilayers and twinning defects were not observed. The quality of the QL structure is readily shown by the juxtaposition of the atoms in the bulk STEM images with a schematic of the ideal structure as seen in the inset. Hexagonal lattice parameters were measured at $a \sim 0.429$ nm and $c \sim 2.958$ nm respectively, which are both within 3% of values found in literature.[35] Aberrations from regular hexagonal shape in the top down images are attributed to drift-induced stretching effects.

HR-XRD characterization also shows a high quality QL structure (Fig. 3). XRD measurements yielded strong diffraction peaks at the (0 0 0 3n) planes, and produced a $c$ value of 2.833 nm, which is within 1% of literature values.[35] The diffraction peak intensities follow those previously reported[8]. Since the measurements of the lattice constants of bulk $Bi_2Se_3$ produced values close to the literature, and top down and cross-sectional lattice images show no signs of defects, the origin of time dependent changes to this crystal sample of $Bi_2Se_3$ presented in the next section are considered to arise from the basic properties of crystalline $Bi_2Se_3$. The quintuple layer (QL) structure and weak vdW inter-QL bonds of the $Bi_2Se_3$ crystal allow for samples to be prepared via micro-mechanical exfoliation, as has been done with graphene since 2004.[5] Scotch tape exposes a fresh QL as the sample surface, enabling measurements on pristine surfaces, and further measurements as the samples age in air.

## 2. AFM Characterization of Time Dependent Morphology

The time dependence of the surface morphology, structure, and chemical state were studies by AFM, ARXPS, XPS, and RA-SH. First, the observation of islands on the surface by AFM is presented. AFM scans completed within 25 minutes from exfoliation show atomically flat surface topographies. Initial surface roughness in the sample shown in Fig. 4 is 0.43 Å RMS. Subsequent scans show the nucleation and growth of patches that within measurement error are uniform in height (3.6 ± 0.2 Å). These patches and their growth process are also noticeably different from samples that experience dirt adhesion or buildup of contaminating particles (see appendix). Their height above background is consistent with the height of Bi bilayers in the bulk of $Bi_2Te_3$[19] which is structurally similar to $Bi_2Se_3$. Based on this consideration, we hypothesize that the patches are bilayers of Bi. The existence of Bi bilayers on the surface of $Bi_2Se_3$ samples is well documented in literature.[8,17] This patchy growth provides an explanation for why surface Bi bilayers are observed by some methods and not by others.[16,17] It also is indicative of a vertical diffusion process, in which Bi rises to the surface. We expect that this diffusion alters the band structure of $Bi_2Se_3$ at the surface, changes the conductance of the topologically protected surface states, and could contribute to 2DEG formation. The majority of patch growth occurs before 2 hours. After 2 hours, the sample appears to roughen ubiquitously, and we attribute this change to surface oxidation in agreement with XPS measurements to follow (Fig. 5**)**. This roughening process slows over time, leaving the background with RMS roughness of 1.1 Å after 22 hours. It is possible that oxide growth up to a certain length of time after the exfoliation provides a barrier for further surface diffusion, thus distinguishing the growth of Bi bilayers on top of the oxidized surface from the initial oxide growth.

## 3. Characterization of Time Dependence of Surface Oxidation by XPS and ARXPS

XPS and ARXPS provide insight into the growth of the oxide overlayer over a period of ~ 1.5 weeks after exfoliating the samples. Fig. 5 contains angle-integrated, time dependent XPS spectra showing aging of the $Bi_2Se_3$ surface, and provides information about the constituent elements by means of the Se 3d, Bi 5d, and O 1s level peaks. Table I lists the values of binding energy of the samples after exfoliation. Spin orbital splitting leads to a pair of peaks for the Se 3d ($3d_{5/2}$ and $3d_{3/2}$; separated by ~ 0.9 eV) and Bi 5d ($5d_{5/2}$ and $5d_{3/2}$; separated by ~ 3.0 eV) levels. The freshly exfoliated sample (t = 0 minutes) shows no signs of oxidation as evident by a flat O 1s region and absence of higher binding energy peaks in the Se 3d and Bi 5d regions. We estimate the binding energy reference for Bi metal $5d_{5/2}$ peak as 24.0 ± 0.3 eV and Se chalcogen $3d_{5/2}$ as 55.4 ± 0.6 eV after calculating mean and average of all the reported values present at the NIST XPS database.[32] Thus, for the freshly exfoliated sample (t = 0 minutes), the Bi $5d_{5/2}$ and Se $3d_{5/2}$ in $Bi_2Se_3$ show a charge shift (CS) of +1.2 eV and –1.7 eV respectively which agrees well with the previously reported values for $Bi_2Se_3$.[13,14] The more electronegative Se (2.55 in Pauling scale) pulls a greater portion of the electron cloud towards itself as compared to the less electronegative Bi (2.02 in Pauling scale), thus pertaining some ionic nature to the bonds in the $Bi_2Se_3$ crystal[36]. No evidence of oxidation is visible in the XPS spectra of samples up to t ~ 119 minutes from exfoliation (Fig. 5). AFM measurements show the growth of small patches (3.6 ± 0.2 Å in height, less than 3 μm in diameter) for t < ~119 minutes, and subsequent roughening of the whole surface. The relative concentration of those patches with respect to the analyzing volume is below the detection limit of our XPS measurement. For samples exposed to ambient air longer than t ~ 189 minutes, a sharp O 1s peak shows up implying adsorption of O on the (0001) surface of the crystal with time. At the same time, new features show up near the Se 3d (broad peak at ~ 59 eV) and Bi 5d (shoulders towards higher binding energy side) regions at higher binding energy values confirming the oxidation of both Bi and Se at the surface. Top layer adsorption would produce O 1s peaks with no corresponding oxide peaks in the case that oxygen did not chemically bond to the surface, or O 1s and Se oxide peaks but no Bi oxide peaks in the case that oxygen only bonded to the topmost layer, as indicated by low patch coverage in AFM. The observed values at the higher binding energy region of the Se 3d and Bi 5d (Table I) are close to the values for $Bi_2O_3$ and $SeO_2$, the most stable oxides of Bi and Se under ambient conditions. As the O 1s peak show only one binding energy value (531.5 eV), we conclude that the average bonding environment around the O atom stays similar over time and the O atom remains surrounded by both Bi and Se atoms, *i.e.*, the chemical environment in the oxide overlayer is similar to a mixture of $Bi_2O_3$ and $SeO_2$. At the beginning of the oxidation process, it is possible that some O-vacancy related defects will remain in the oxide overlayer. Adsorption of more oxygen atoms over time will heal some of these defects and also increase the thickness of the oxide overlayer as more Bi and Se from the QL layer reacts with the atmospheric O. The O 1s, Se 3d oxide, and Bi 5d oxide peaks all continue to grow as evidenced by spectra collected at subsequent times (Fig. 5) although the rate of growth of the oxide overlayer appears to decrease after ~8 hours. This may be due to the vdW gapped structure of $Bi_2Se_3$, which could act as a temporary barrier to oxidation beyond the first QL.[25] We expect that over longer time scales (months), oxidation may continue through further QLs via defect-mediated movement of O in the intercalation layer. The time dependence of the oxidation process is quite unusual, in that it displays a significant delay between surface exposure and oxygen incorporation. It has been reported that a delay to oxidation of up to 20 hours is normal for the (0001) surface of certain $Bi_2Te_3$ crystals.[25] Further study is required relating the common factors between these two crystals (e.g., Bi presence, TI phase, QL structure, etc.) to the delayed oxidation behavior.

Analysis of angle-resolved XPS data provides a profile of the growth of the oxygen overlayer with time. Among other factors, photoelectron counts are a function of collection angle from normal incidence ($\theta$), and electron origin depth (inset of Fig. 7(a)). This is largely due to inelastic scattering events in the sample[37], which can be approximately modeled with a Beer-Lambert function: $I = I_o * \exp[-\alpha x/\{d\cos(\theta)\}]$, where $I_o$ is the intensity of emitted electrons, $d$ is the vertical depth from the sample surface to the origin of the emitted electron, and $\alpha$ is a material-dependent scatterer density parameter. Fig. 6 shows the representative plots for Se 3d and Bi 5d regions' ARXPS data collected for the sample ~1.5 week old. The relative intensities of the oxide signals increase in going from a smaller angle (more sensitive to signals coming from the bulk) to angles close to the surface implying the depth of the oxide overlayer is limited to few QLs during our measurement time. Applying a maximum entropy approach[38,39] to the normalized ARXPS data as implemented in the ARProcess software in Thermo Avantage, growth of the oxide overlayer was quantified as shown in Fig. 7. Oxide depth appears to increase from 0 nm before 119 minutes to 1 nm at 189 minutes, to ~1.9 nm at the end of 1.5 weeks. This matches previously reported values very closely, which describe the oxidation process of the top-most QL in $Bi_2Se_3$.[15] It also indicates that the expansion of a QL due to oxidation increases its volume by ~2x from 0.96 nm in an oxidized QL. This expansion rate is similar to that seen in Si[40].

## 4. VASE Characterization of the Time Dependence of Surface Oxidation

Variable angle spectroscopic ellipsometry was used to characterize the changes in freshly exfoliated $Bi_2Se_3$ continuously over a 27-hour period. Three repetitions of the experiment, each at different angles, gave consistent results. Standard ellipsometry data $\Psi$, the relative polarization amplitude change, and $\Delta$, the relative phase shift of polarizations, are shown in Figs. 8(a, b). After exfoliation, for the first three hours, $\Delta$ decreases rapidly and then slowly decreases, while $\Psi$ monotonically increases over time (Fig. 8(c)). Changes in $\Psi$ are small at most energies (wavelengths) due to thinness of the oxide layer. However, $\Delta$ is sensitive to small changes in thin dielectric films, including most oxides. The well-known Drude approximation indicates that $\Delta$ is a linear function of film thickness for thin dielectric films on metals or silicon.[41] Here, $\Delta$ shows significant change at higher energies (smaller wavelengths) over the time frame when surface patch growth is observed by AFM. Because XPS does not observe an oxide layer prior to 3 hours, the changes in $\Delta$ coincide with formation and growth of surface patches. Sample thicknesses of > 1 mm lead to complete absorption of light above the 0.35 eV bandgap. Thus, bulk $Bi_2Se_3$ behaves as a substrate for our entire measurement range of 1.0 to 4.0 eV, and is treated accordingly in our models. Initial measurements after exfoliation are used to make a simple substrate model and extract a dielectric function of bulk $Bi_2Se_3$. As can be seen in Fig. 8(d), the main peak in the imaginary portion of the dielectric function, $\varepsilon_2$, is located at 2.00 eV, and the real portion of the dielectric function, $\varepsilon_1$, crosses zero at 2.27 eV. Utilizing these constant substrate properties and the observed changes in $\Psi$ and $\Delta$ over time, we are able to extract the dielectric function of the surface oxide. The oxide's dielectric model is composed of Tauc-Lorenz oscillators, and has a band gap of 2.7 eV, which matches previously reported values for Bi oxide.[42,43] Se oxide may also contribute to $\varepsilon_2$ at energies above its band gap of 3.5 eV[44], but we cannot determine either oxide's individual dielectric function from the single oxide layer model in use here. Additionally, due to the dynamic changes in the surface immediately after cleaving (Se vacancy formation, possible Bi diffusion, 2DEG formation), we consider this oxide growth model to be more representative of the observations from the other data following the third hour after exfoliation. In that regime, time dependence of the oxide thickness corresponds to the change in Ψ and Δ. In addition to providing information about electronic behavior, these dielectric functions (refractive indices) can be used to model reflectivity and the contrast of $Bi_2Se_3$ flakes on variety of substrates as seen in an optical microscope with the Fresnel equations at normal incidence:

$$r_{0,1} = \frac{n_0 - n_1}{n_0 + n_1} \qquad \beta_k = \frac{2\pi n_k d}{\lambda} \qquad (1)$$

$$r_{0-k+1} = \frac{r_{0-k} + r_{k,k+1}\exp(-i2\beta_k)}{1 + r_{0-k}r_{k,k+1}\exp(-i2\beta_k)}$$

Here, $r_{0,1}$ represents the electric field reflectance at the interface between layers *0* (taken as ambient) and *1*, $n_k$ is layer k's complex refractive index, and $\beta_k$ gives the absorption of and phase acquired by light of wavelength $\lambda$ propagating through layer $k$ of thickness $d$, and $r_{0-k+1}$ is the field reflectance produced by layers *0* through *k+1* The reflection intensity is given by $|r|^2$. Fig. 9 shows maps of the contrast created by unoxidized flakes on a Si wafer, as well as on a Si wafer with 300 nm of thermal oxide, as a function of flake thickness and incident wavelength. For flakes between 0 and 20 nm thick, an increase in thickness causes an increase in contrast with the background. However, since the trend across wavelengths changes minimally with changes in thickness, the change in color of a flake may not be a reliable metric for flake thickness.

## 5. Nonlinear Optical Processes using Second Harmonic Generation

Second harmonic generation (SHG) is a non-linear optical technique that probes for surface-specific electronic properties and the lattice symmetry of $Bi_2Se_3$. The generated second harmonic light of frequency 2ω using the incident fundamental light at frequency ω is forbidden in centrosymmetric systems such as $Bi_2Se_3$. However, at the surface, there is a break in the symmetry due to the transition to ambient, allowing SHG from the crystal surface. In our rotational-anisotropy SHG (RA-SHG) experiment, the sample was rotated azimuthally about the surface normal (0001), and azimuthal angle dependent SHG intensity was measured for p-in/p-out polarization configuration. In Fig. 10, one can see that the 3-fold symmetric RA-SHG signal appears at the beginning stage after sample cleaving, corresponding to the 3-fold symmetry of the $Bi_2Se_3$ crystal, and it develops over time. Initially, there are three major lobes at 60°, 180°, and 300° and three minor lobes at 0°, 120°, and 240°, of sample azimuthal angles. The minor lobes appear in a fresh sample when the outgoing laser is nearly aligned with the (0 15) plane normal direction, and the major lobes appear in a fresh sample when the incident laser is nearly aligned with the (01) plane normal direction, as verified with XRD (Fig. 11). As the (0 15) plane normal is parallel to the Bi-$Se_{(2)}$ bond at the top QL surface, alignment between the top Bi-$Se_{(2)}$ bond and the outgoing laser produces minor lobes, and alignment between the top Bi-$Se_{(2)}$ bond and the incident laser produces major lobes. The major lobes decrease in amplitude while the minor lobes increase, eventually becoming equal in amplitude for a 6-fold symmetric RA-SHG scan at ~3.5 hours. Subsequently, the lobes continue to grow at 0° (and other equivalent angles) and shrink at 180°, thus switching their major and minor characters. When the sample has not experienced any prior laser exposure history, this major-to-minor lobe switching process is complete at 12 hours with a majority of the change occurring by 6 hours. Analysis was done with the following SHG field equation for light that was p-polarized before and after reflection:

$$E^{2\omega}(\phi) = a_0 + a_3\cos(3\phi) \qquad (2)$$

Here $a_0$ and $a_3$ are respectively the isotropic and anisotropic complex coefficients, both determined by Fresnel factors of the beam configuration and dielectric susceptibility of $Bi_2Se_3$, and $\phi$ is the sample azimuthal angle between the incident plane and the mirror symmetry plane, with the directionality defined as in Fig. 10(d). The SHG intensity measured is proportional to . Both coefficients are time-dependent, written as $a_0(t)$ and $a_3(t)$, because the oxide/$Bi_2Se_3$ interface translates inwards inside a QL. Figure 10 shows the measured rotational-anisotropy SHG (RA-SHG) scans in a dry air environment immediately, and at 1, 2, 3, 4, 12, and 24 hours after $Bi_2Se_3$ sample cleaving. To avoid prior laser exposure, the sample was shifted to a different measurement spot before each RA-SHG measurement. The thin lines are experimental data of SHG, and the thick smooth curves are fits to Eq. (2). The $a_0(t)$ term contains a contributions from the surface dipole induced SHG and electric field induced SHG. The relative contribution of these contributions change as the surface oxidizes. Fig. 10(b) shows empirical fits to $a_0$ and $a_3$ over time. The major change to $a_3$ to occurs within 1 hour. As other measurement methods detect no oxide in that time, we attribute this change in $a_3$ to be either caused by Se vacancy formation or 2DEG formation from band bending. After the first hour, the value of $a_3$ remains near 1.44. The effects of charge accumulation, however, change the isotropic parameter $a_0$ from -0.19 to +0.19 while the first QL is being oxidized. The $a_0$trend closely resembles the trend of oxide thickness observed in the previous sections of this work. We thus consider $a_0$ to be a valid metric for measuring oxide thickness via RA-SHG.

We can reasonably assume that a 6-fold RA-SHG pattern corresponds to the time when the top two layers and part of the middle layer of the top QL is oxidized, and a full major-to-minor lobe switching corresponds to the time when a full QL is oxidized. In comparison to the XPS results with regard to the time scales of surface oxidation, RA-SHG shows a gradual evolution of oxidation within a QL during the 12 hours period after sample cleaving.

### 6. Description of Time Dependent changes in $Bi_2Se_3$ (0001) Surface

The surface aging process occurs in sequential stages. First, surface islands of height 3.6 ± 0.2 Å form over the first 25 minutes as measured by AFM. The ellipsometry parameter Δ also changes during the same time period. Since no O 1s peak is observed by XPS, the observations are interpreted as being due to segregation of Bi to the surface over the first 25 minutes. The AFM data shows that patch growth continues for the first two hours, followed by surface roughening. The presence of a surface oxide layer composed of both Bi oxide and Se oxide is clearly present after 189 minutes (~3 hours) in the XPS data. Calculated values of the oxide thickness from both XPS and SE find that the oxide film present at ~3 hours is ~1 nm thick. Using the estimated expansion factor of 2 (as discussed in section **3**), the first ½ QL is oxidized at ~3 hours. Here we refer to the oxidation of the top two layers plus part of the middle layer as the top ½ QL. The RA-SHG data show that half of the change in the top QL has occurred by ~3 hours, based on the six-fold symmetric RA-SHG pattern and the associated data from other methods. Thus the oxidation starts soon after 2 hours and continue until the first QL is oxidized at ~1.5 weeks. The RA-SHG for the p-in/p-out polarization combination shows that most of the oxidation induced change is done after ~12 hours. There is only a minimal change in the RA-SHG data between 12 and 24 hours. The single oxide film model for the SE data also shows a small change in oxide thickness between 12 and 24 hours.

## IV. CONCLUSION

We present a unified understanding of the changes that occur at the (0001) surface of high quality $Bi_2Se_3$ when exposed to air based on surface characterization and electron microscopy. AFM measurements provided the first evidence for the growth of small Bi bilayer patches. This explains the inconsistency of their presence in various previous microscopic measurements. The time-dependent topography also provided information about oxide growth and corresponding roughness. The incubation time for oxidation of the top QL with an oxide thickness of 1.9 nm after 1.5 weeks from exfoliation were determined using the Bi 5d, Se 3d, and O 1s ARXPS spectra. The findings that $Bi_2Se_3$ does not immediately oxidize in air contrasted some previous XPS findings.[22] We confirmed the oxide growth dynamics and depth, as well as measured the dielectric functions of bulk $Bi_2Se_3$ and its surface oxide with VASE. We find that Δ was sensitive to the patch growth observed by AFM. These were used to produce contrast maps from the Fresnel reflection equations, which provide a key to optically identify $Bi_2Se_3$ flakes by their color and thickness. Finally, the results of the time dependent ARXPS, SE, and AFM characterization were used to more completely interpret time dependent RA-SH of the $Bi_2Se_3$ (0001) surface in terms of progression through the top QL structure, and develop a highly effective oxide depth measurement method. The results of this study provide a more complete understanding of the impact of oxidation on the topologically protected states on (0001) $Bi_2Se_3$.

## 7. ACKNOWLEDGEMENTS

We acknowledge the SRC NRI INDEX center, which provided partial funding for this study. We would also like to give special thanks to Dr. Ken Burch, who has provided invaluable discussion about the field of TI materials. We also acknowledge Doug Medlin for discussions about the microscopy of $Bi_2Se_3$. We would like to thank Dr. Michael Hatzistergos of the Colleges of Nanoscale Science and Engineering at the SUNY Polytechnic Institute for technical assistance and discussions regarding the XPS measurements.

## 8. APPENDIX

Two topics of import to this paper involved the distinction between surface patches and dirt as seen in AFM, and the distinction between a shake-up peak and an oxide peak in the Bi 4f energy range of XPS.

### 1. AFM Topography Verification

Fig. 12 shows the topology of dirt, which creates stark contrast to the consistent and thin patches that grow from the crystal surface. Thus, we are assured that the thin patches we observe are not due to the deposition of dirt or other physical contaminants. Rather, they indicate a change in surface morphology due to changes in surface chemistry and structure.

### 2. Bi 4f Shake-Up Exposition

Fig. 13 shows the time evolution of the Bi 4f XPS peaks. At t = 0 minutes, the main Bi $4f_{7/2}$ and $4f_{5/2}$ XPS peaks are accompanied by shoulders to the higher binding energy side. Although the Se 3p 3/2, 1/2 doublet and shake-up alter the higher binding energy spectrum of the Bi 4f peaks, the effect of oxidation can be observed As there is no evidence of oxygen adsorption at the (0001) surface for $0 < t <$

~ 119 minutes (see the O 1s spectra in Fig. 5c), the presence of these shoulders towards higher binding energy sides in the Bi 4f XPS region may not be used for interpreting oxidation of the (0001) surface of $Bi_2Se_3$. These shoulder peaks are attributed to various shake-up events[23,24] and possible overlap with peaks from Se[45] in the literature. For samples exposed to the ambient air for longer time periods, the shoulder peaks grow in intensity as the contribution from the Bi 4f oxide peaks also appear in the same region. The Bi 5d peak region (Fig. 5b) is devoid of any such complications arising from peak overlap. Thus, we used the Bi 5d XPS peak region, along with the Se 3d and O 1s regions, to understand the growth mechanism of surface oxide on the $Bi_2Se_3$ (0001) surface.

## 3. SHG Orientation Verification

Fig. 14 shows a reciprocal space map (RSM) of a (0 1 $\overline{1}$ 5) peak. It is one of the three-fold symmetric peaks that, due to the structure factor of $Bi_2Se_3$, allowed us to determine the crystal's orientation. With this data and clear substrate alignment markers, we were able to determine the relationship between SHG peaks and bond orientations in the $Bi_2Se_3$ unit cell.

TABLE I. Results of time dependent XPS and ARXPS investigation on $Bi_2Se_3$ after exfoliation along (0001).

| Sample id (time from exfoliation) | Binding energy (eV) ± 0.05 eV | | | | | Thickness (nm)± 0.2 |
|---|---|---|---|---|---|---|
| | Bi $5d_{5/2}$ in | | Se $3d_{5/2}$ in | | O 1s in | |
| | $Bi_2Se_3$ | Oxide over-layer | $Bi_2Se_3$ | Oxide over-layer | Oxide over-layer | |
| t ~ 0 min | 25.2 | -- | 53.7 | -- | -- | -- |
| t ~ 119 min | 25.2 | -- | 53.7 | -- | -- | -- |
| t ~ 189 min | 25.2 | 26.4 | 53.7 | 58.7 | 531.5 | 1.0 |
| t ~ 309 min | 25.2 | 26.4 | 53.7 | 58.7 | 531.5 | 1.4 |
| t ~ 479 min | 25.2 | 26.4 | 53.7 | 58.7 | 531.5 | 1.6 |
| t ~ 1.5 week | 25.2 | 26.4 | 53.7 | 58.7 | 531.5 | 1.9 |

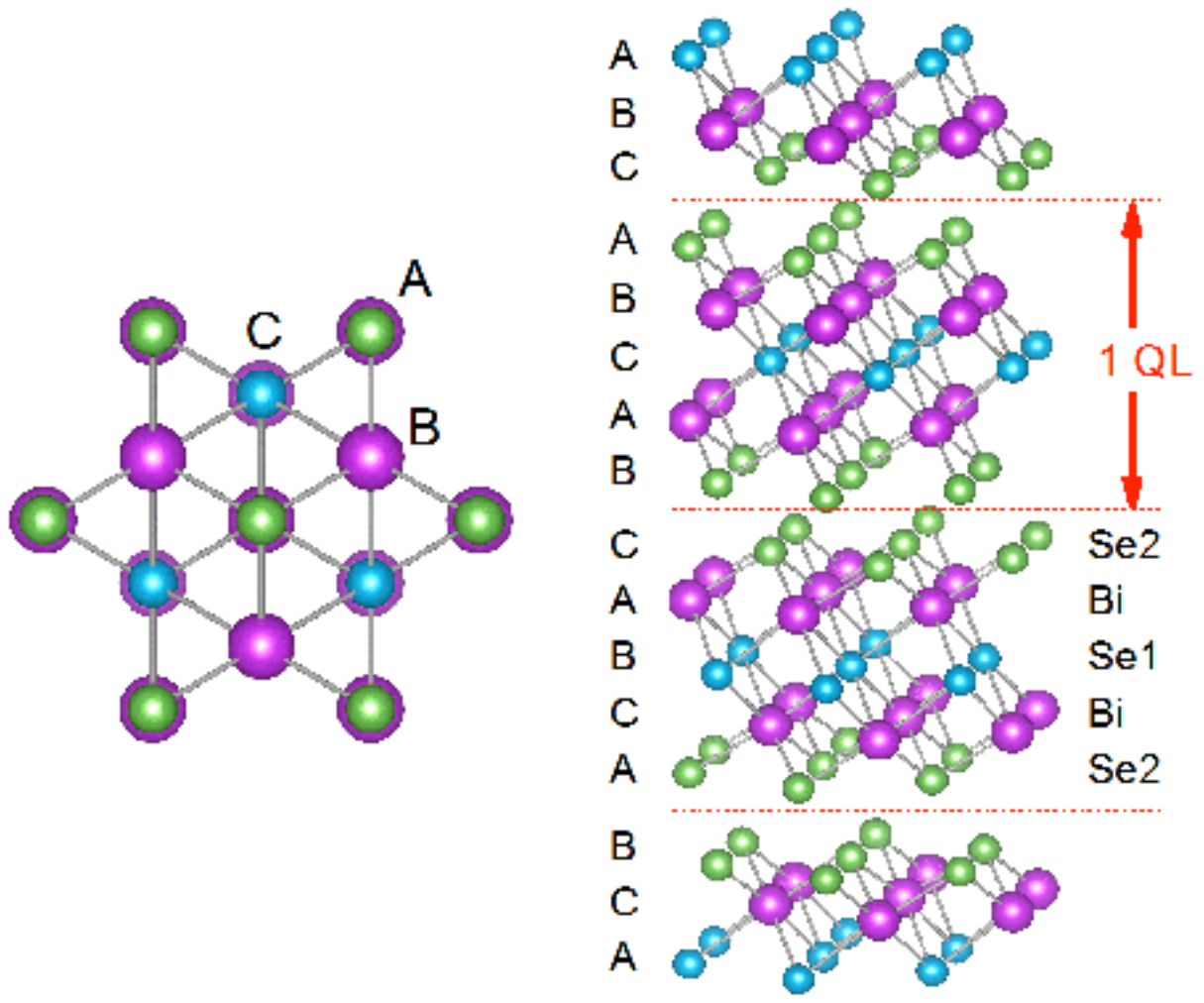

FIG. 1. The layered structure of $Bi_2Se_3$. A unit cell of $Bi_2Se_3$ is composed of three quintuple layers (QLs) where each QL consists of five rhombohedrally stacked atomic layers of Se-Bi-Se-Bi-Se that occupy A, B, and C sites. The QLs are bound by a weak van der Waals interaction making them easy to cleave at the (0001) face. All atoms have octahedral nearest neighbor arrangement. The Se1 atoms are surrounded by six equidistant Bi atoms, and the Se2 atoms are surrounded by three nearest neighbor Bi atoms situated in the same QL and three Se2 atoms belonging to the adjacent QL. The unit cell has lattice parameters $a \sim 0.42$ nm (measured between neighbor sites in the same vertically oriented plane) and $c \sim 2.86$ nm where each QL is $c/3 \sim 1$ nm thick[35].

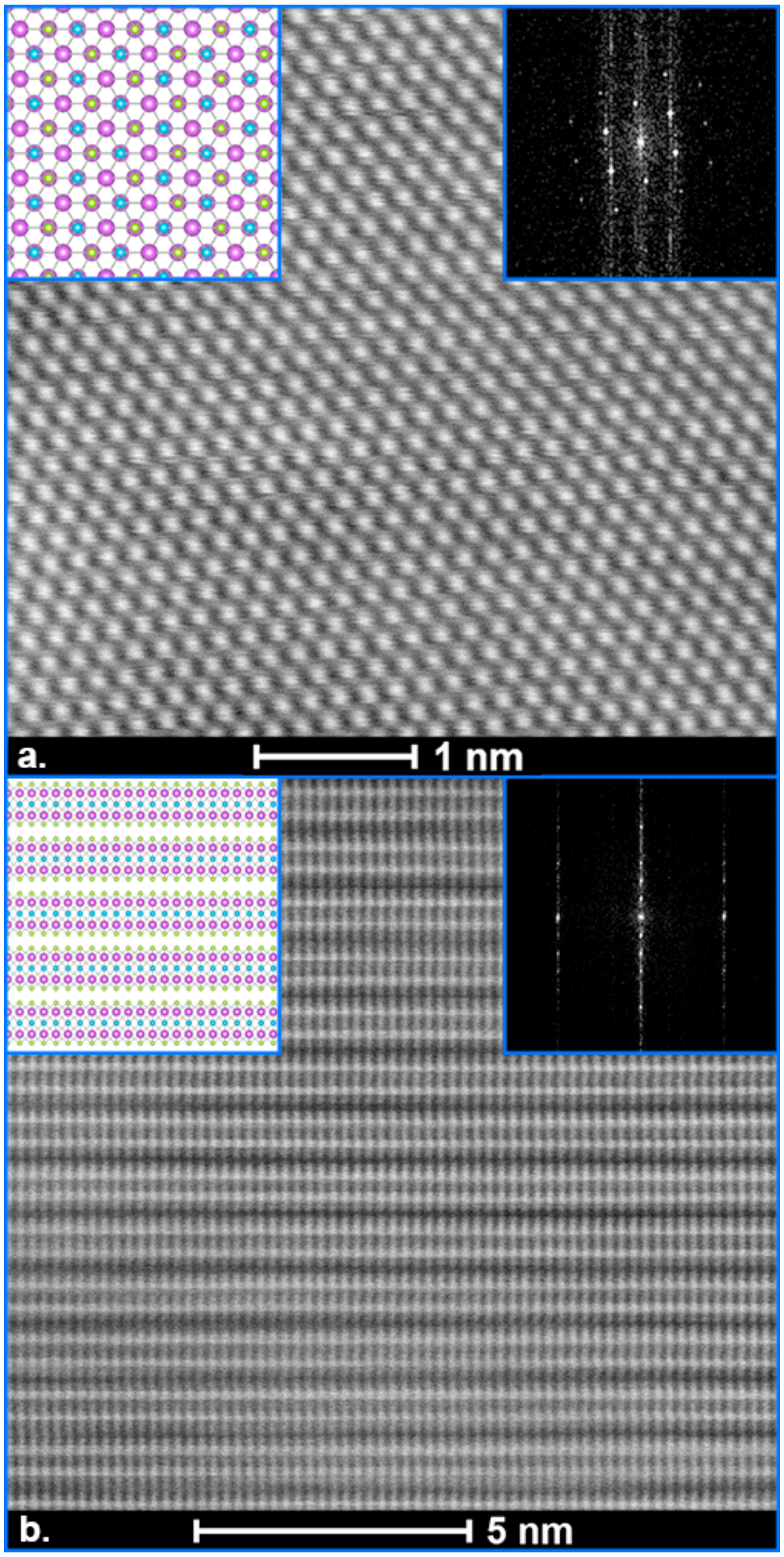

FIG. 2. Aberration-corrected HAADF STEM images of bulk $Bi_2Se_3$. (a) is a top-down image of the crystal along the [0001] zone axis, showing its hexagonal structure. Though each QL has a honeycomb structure, their rhombohedral stacking leads to a 2D hexagonal close-packed image. A hexagonal unit cell is outlined in red. Aberrations from a regular hexagonal shape are attributed to sample drift; (b) is a cross-sectional image of the crystal. In hexagonal notation, the zone axis is normal to the plane $(0\ 1\ \bar{1}\ 0)$. The QL structure of the crystal is apparent from the vdW gaps separating sections of five atomic layers. One can also see the Se-Bi-Se-Bi-Se order of QLs from the relatively high brightness of the Bi atoms. A QL is outlined in green, with the red dotted extension outlining a unit cell. Each figure has an ideal atomic schematic and a power spectrum in the insets. The lattice parameters measured from these images are $a$ = 0.429 nm and $c$ = 2.958 nm.

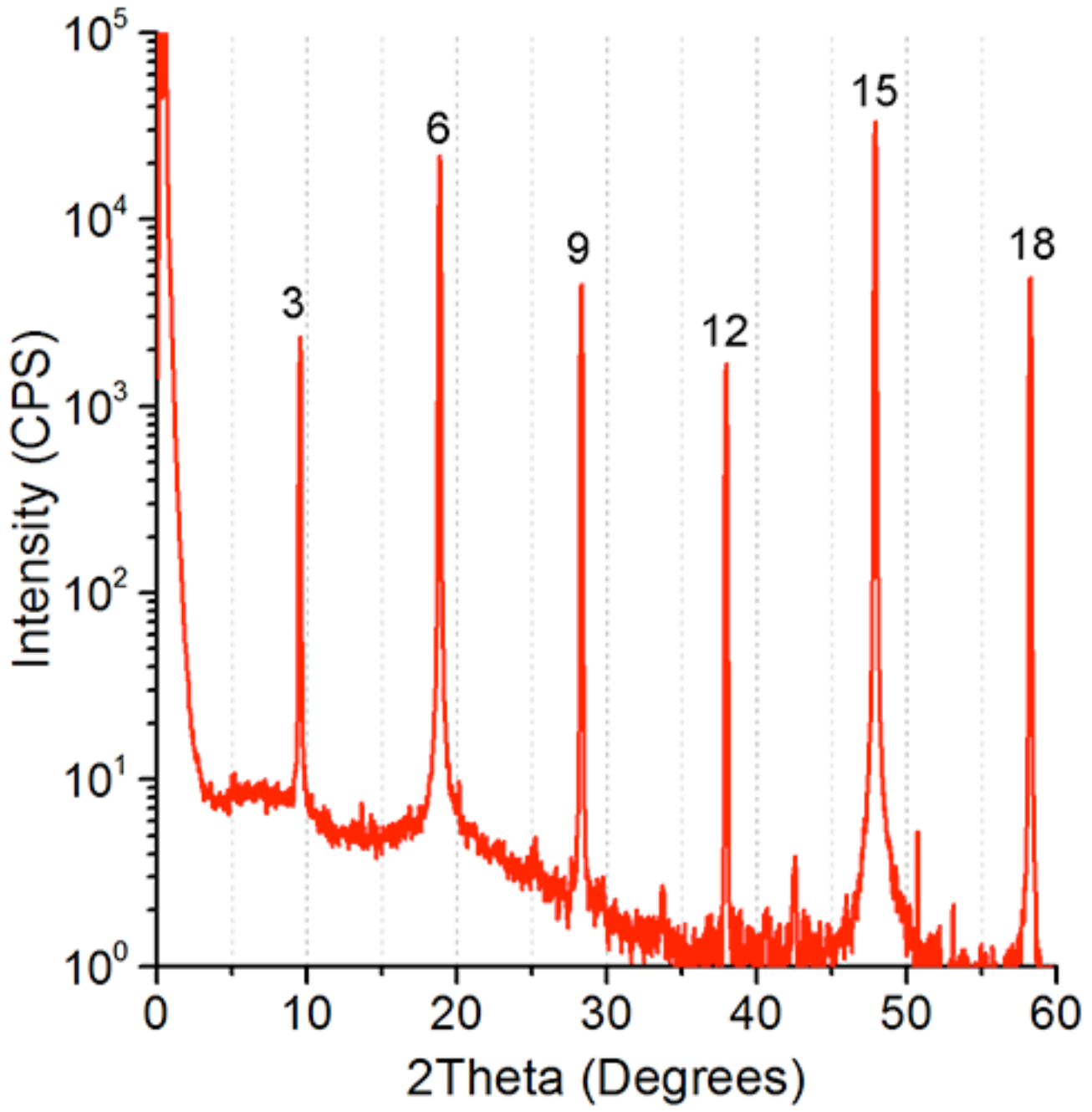

FIG. 3. Omega-2Theta XRD scan of $Bi_2Se_3$. This measurement was performed on a large granule, of approximately 2x2 cm in area and 0.5 mm in thickness. The (0 0 0 3n) planes are the only c-axis planes that are not forbidden. Thus, we have labeled each peak with its k index. These scans give values of 28.427 Å per unit cell, or 9.476 Å per QL.

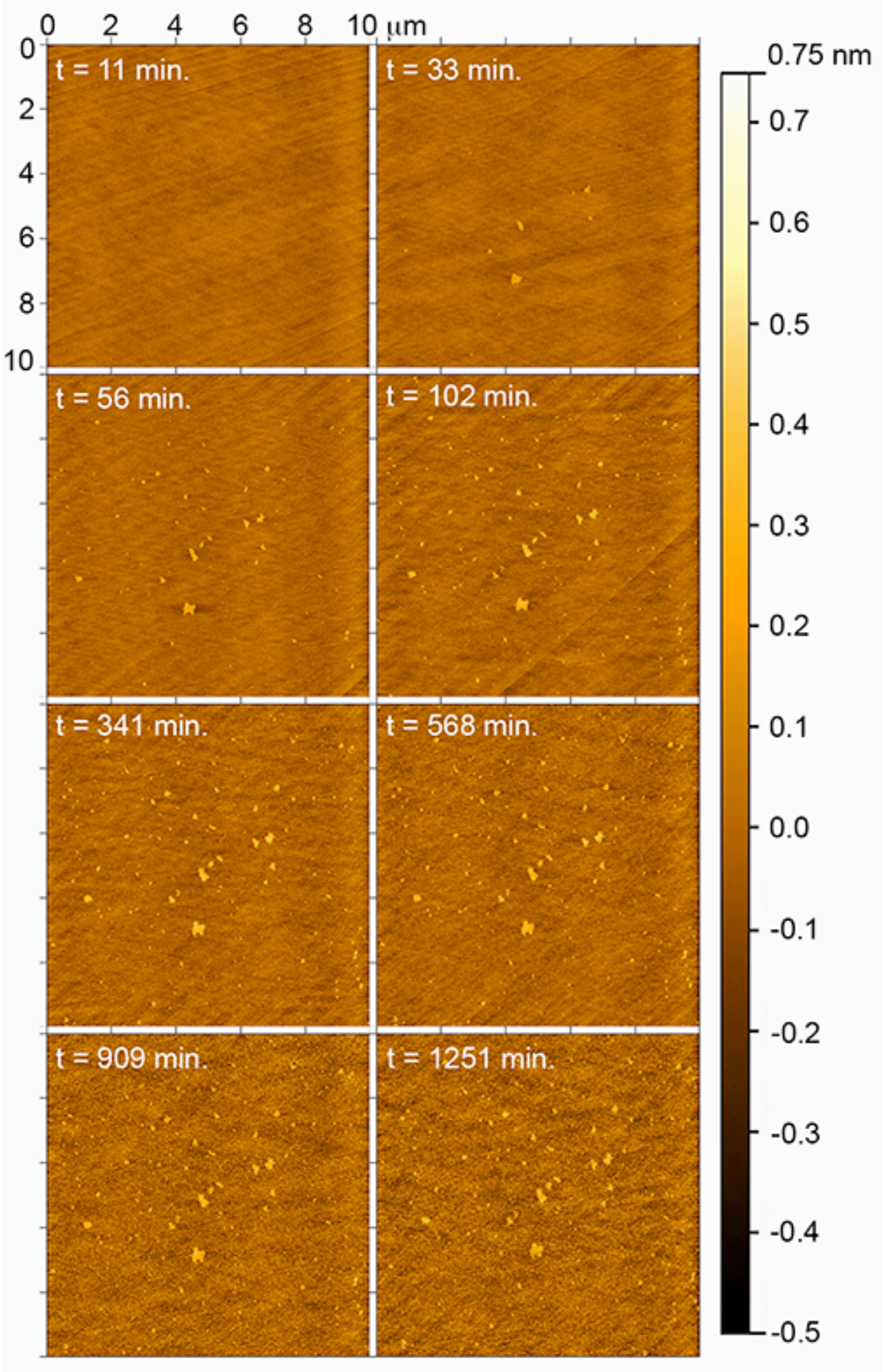

FIG. 4. Time-dependent surface topography of $Bi_2Se_3$. All scans were taken with Bruker TESPA AFM tips. Until 33 minutes after exfoliation, the crystal is atomically smooth. After 33 minutes, surface patches arise, but stop growing at ~100 minutes. They are very uniform in height (3.6 ± 0.2 Å). We hypothesize that these patches are Bi bilayers. The background roughens through the full time scale, increasing from 0.4 Å RMS at 11 minutes to 1.1 Å RMS at 1251 minutes.

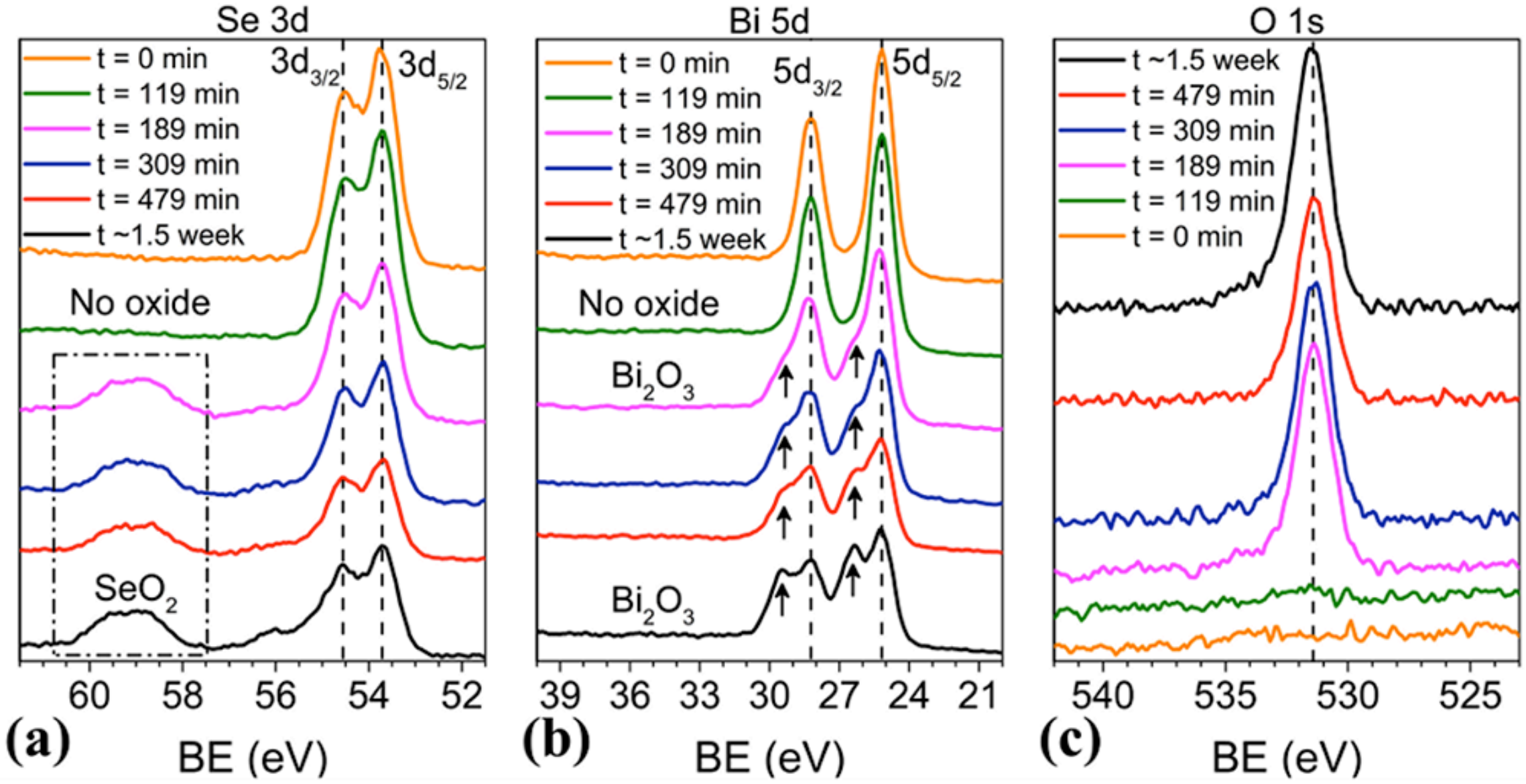

FIG. 5. XPS and ARXPS results. Angle summed XPS spectra of (a) Se 3d, (b) Bi 5d, (c) O 1s region for various times after exfoliation. The oxide peaks corresponding to $SeO_2$ (marked by a dotted box in (a)) and $Bi_2O_3$ (marked by vertical arrows in (b)) are observed after ~119 minutes from exfoliation along with the observation of a strong O 1s peak. The results also indicate that oxygen adhesion and subsurface oxidation proceed simultaneously.

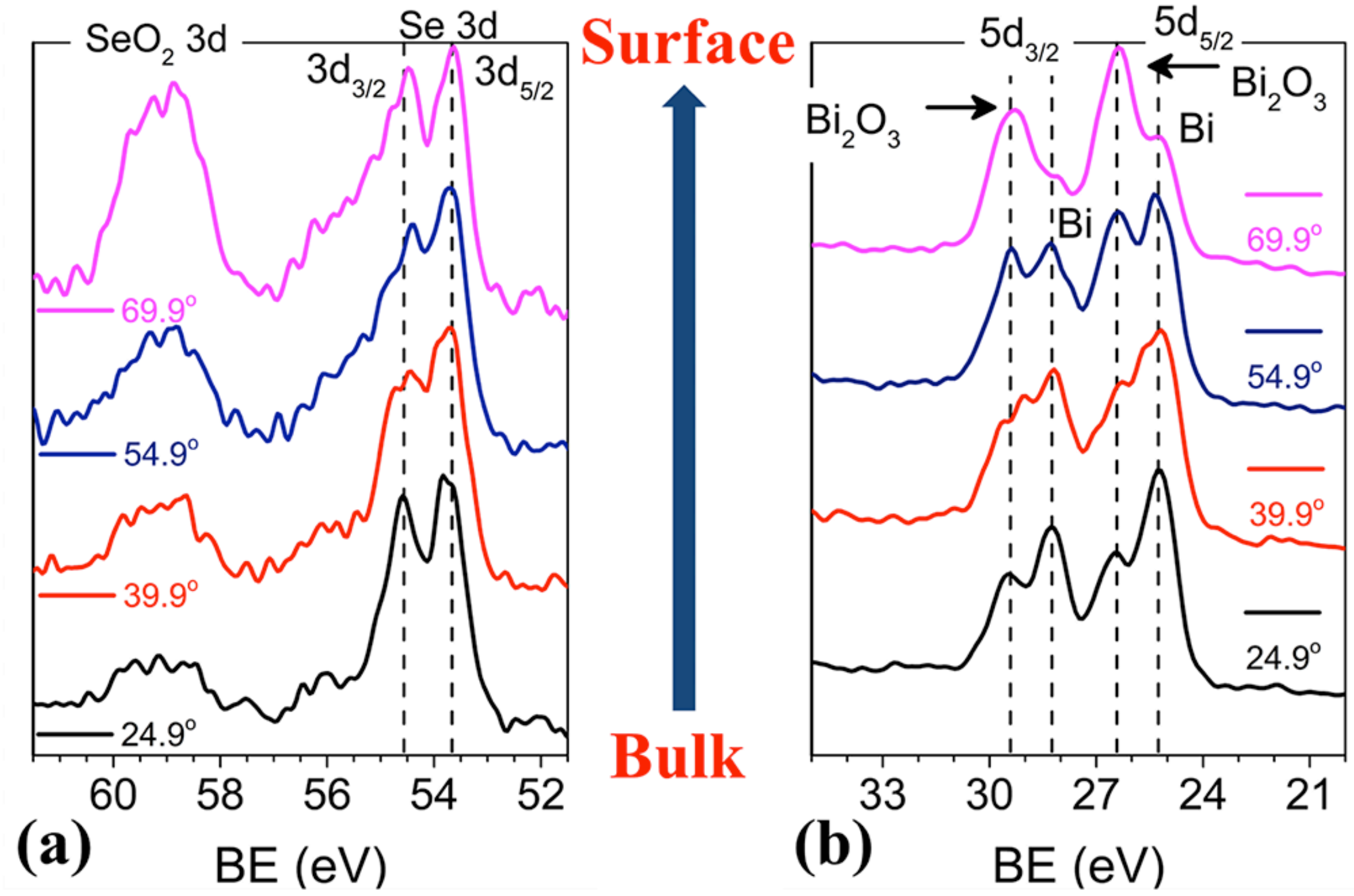

FIG. 6. ARXPS spectra of (a) Se 3d and (b) Bi 5d region for the sample after ~1.5 weeks of exfoliation. The angle in ARXPS profile is measured from the surface normal. Spectra at smaller angles are from layers deeper into the material (bulk) and the spectra at higher angles are coming from layers closer to the top surface of the sample. The relative intensities of the signals due to oxidized Bi and Se increases for spectra collected from layers closer to the surface (24.9° compared to 69.9°) indicating the overlayer is rich in oxygen.

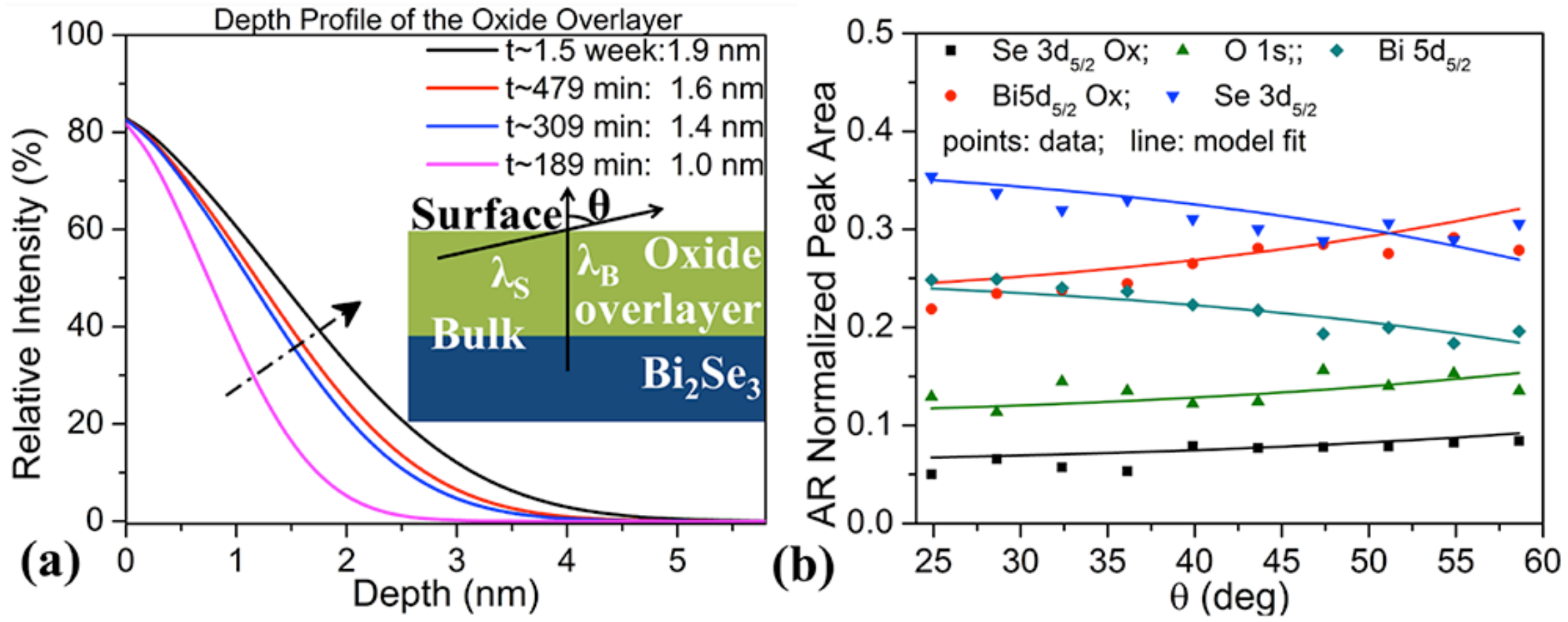

FIG. 7. ARXPS results. (a) Relative intensity depth profile for the oxide overlayer for various times after exfoliation (t ~ 189 minutes to ~ 1.5 week). The thickness of the oxide layer is estimated from these plots where the relative intensity profile comes down to 50% of the maximum value. The oxide layer grows over time, and appears to reach a metastable equilibrium at ~ 1.9 nm after ~ 1.5 weeks of exfoliation (values mentioned in the plot legends). The figure in the inset shows that the information depth varies as a function of the angle (θ) with respect to the surface normal. A smaller value of $\theta$ corresponds to signals coming from layers that lie deep in the material (bulk) and a higher value is more sensitive to the features close to the surface. The Relative intensity depth profile are obtained after modeling the angle resolved normalized peak area as shown in (b) for the t ~ 1.5 week sample. The ARXPS modeling is done according to the maximum entropy approach[38,39]. The peak areas are calculated from the plots in Fig. 6 and normalized with the instrument transmission function, sensitivity factors and energy compensation factor. The calculations are limited between 25° to 60° to minimize the effect of elastic scattering at higher angles[37].

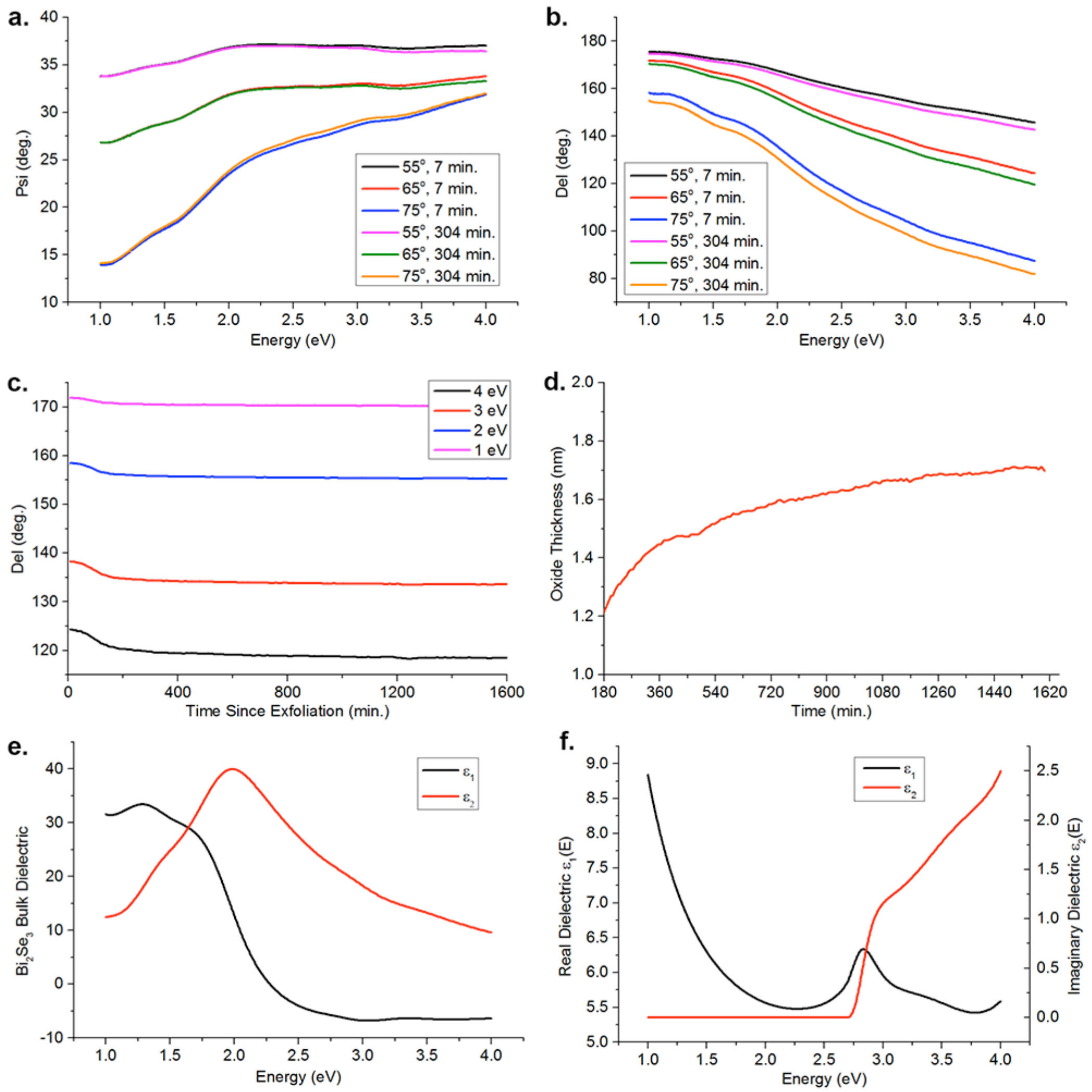

FIG. 8. Variable angle spectroscopic ellipsometry data and results of optical modeling. (a) and (b) show Ψ(E) and Δ(E) at 7 minutes and ~5 hours after exfoliation, at angles of incidence between 55 and 75 degrees from the surface normal. (c) shows Δ(t) for various energies, and highlights the rapid initial decline of Δ at higher energies. (d) shows the growth of oxide on $Bi_2Se_3$ over time. The rapid initial growth of oxide corresponds to the initial decline of Δ. Figures (e) and (f) show the real and imaginary dielectric functions of bulk $Bi_2Se_3$ and $Bi_2Se_3$ oxide, respectively.

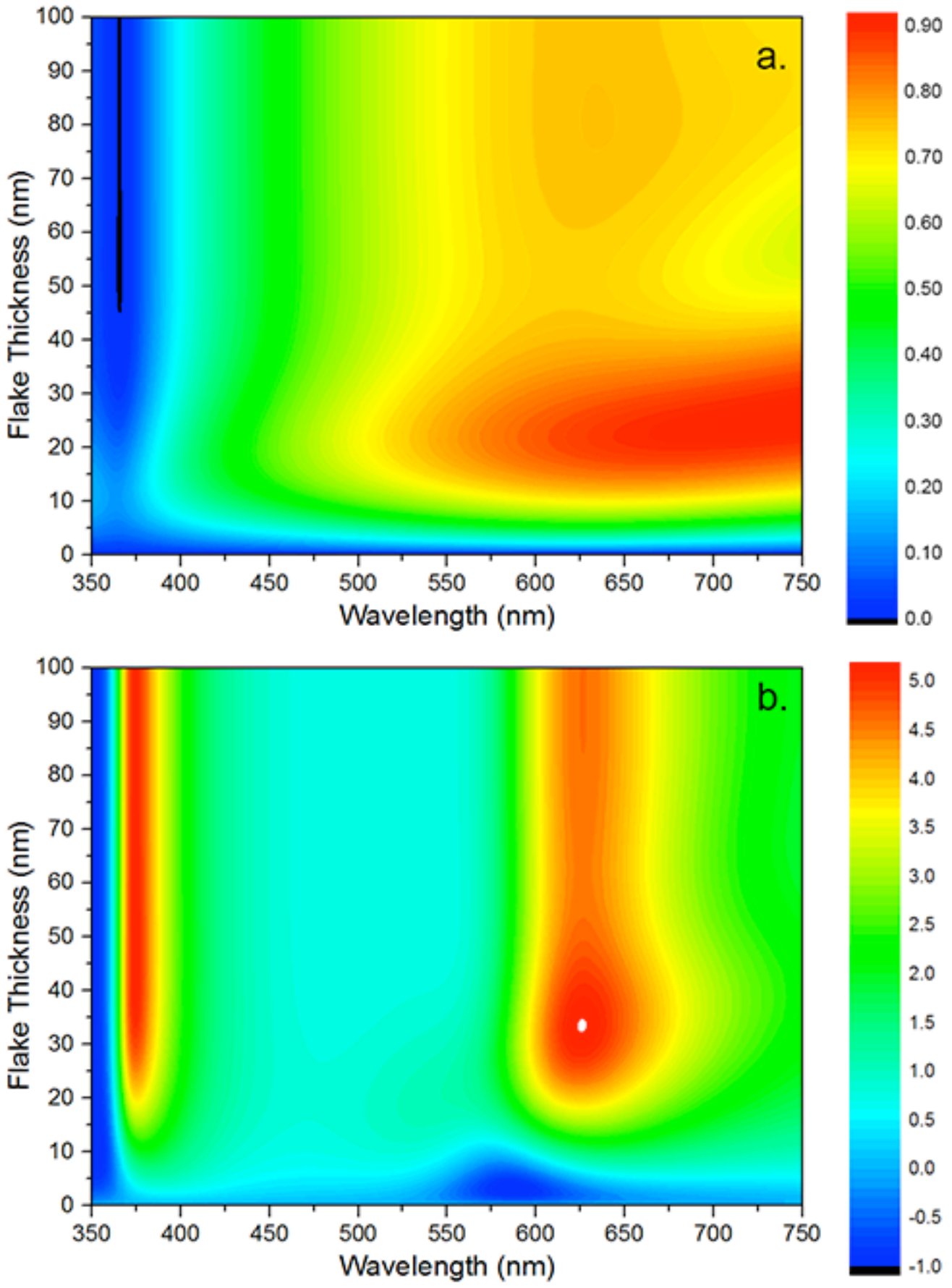

FIG. 9. Contrast maps for $Bi_2Se_3$ on a plain Si substrate (a) and on a Si substrate with 300 nm thermal $SiO_2$ (b). Red regions indicate higher reflectivity of a flake at a given wavelength and flake thickness relative to the reflectivity of the substrate. Likewise, blue regions indicate lower reflectivity of a flake relative to the reflectivity of the substrate. For regions at 0 contrast, there is no difference between the intensity of light reflected from the flake and from the substrate. It is clear that having a 300 nm thermal $SiO_2$ layer between the substrate and the flake causes higher levels of contrast, especially in the d375 nm and 630 nm regions.

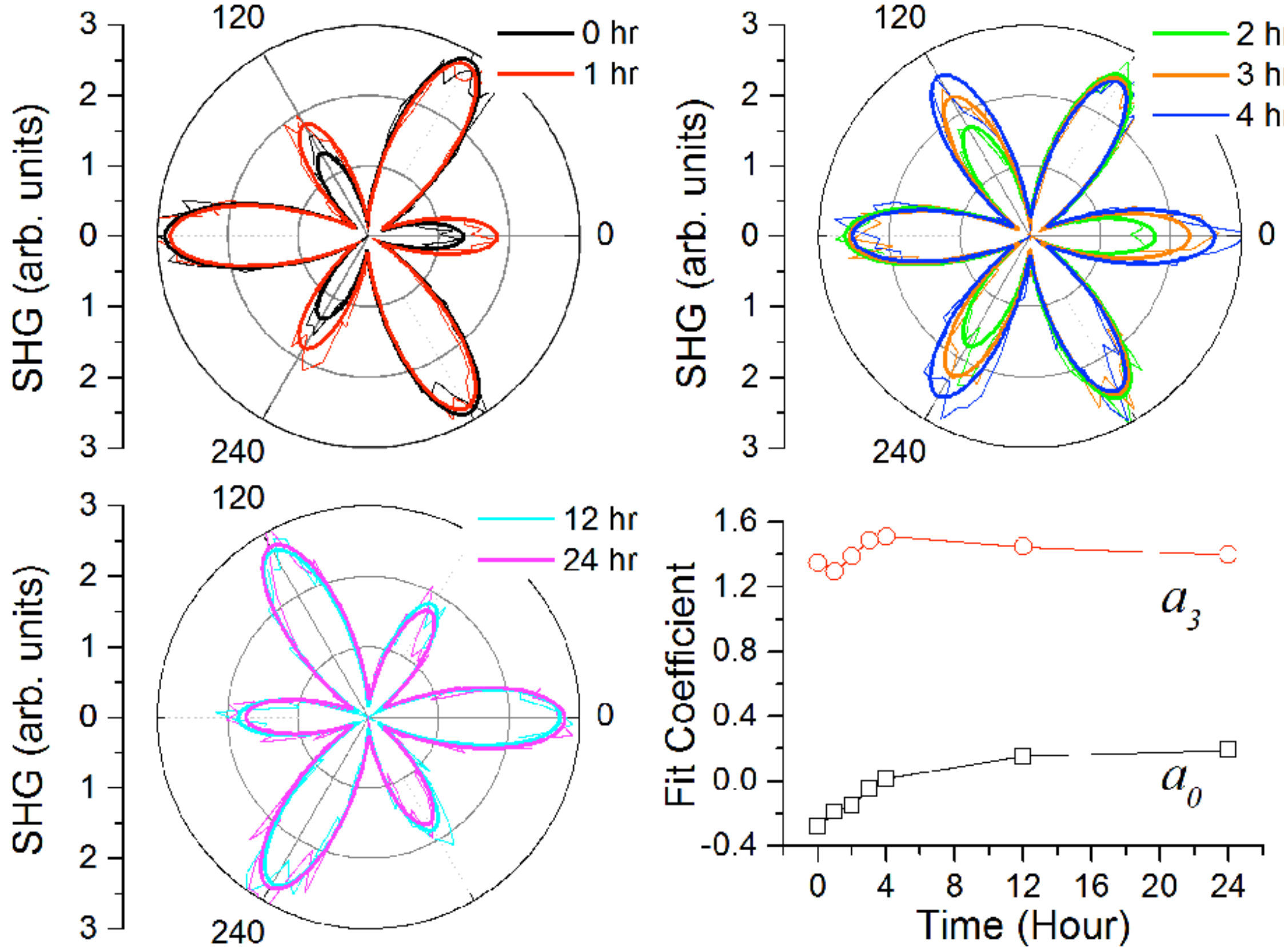

FIG. 10. Measured rotational-anisotropy SHG (RA-SHG) scans in a dry air environment immediately, and at 1, 2, 3, 4, 12, and 24 hours after $Bi_2Se_3$ sample cleaving. Each scan was taken at a different sample spot so that the measurement spot had not experienced any prior laser exposure before each measurement. Thin lines are experimental data, and thick smooth curves are fits to Eq. (2). The empirical fit parameters $a_0$ and $a_3$ are plotted as a function of time. While $a_3$ only undergoes minor changes in the early stages of measurement, $a_0$ increases significantly and monotonically throughout the full oxidation process.

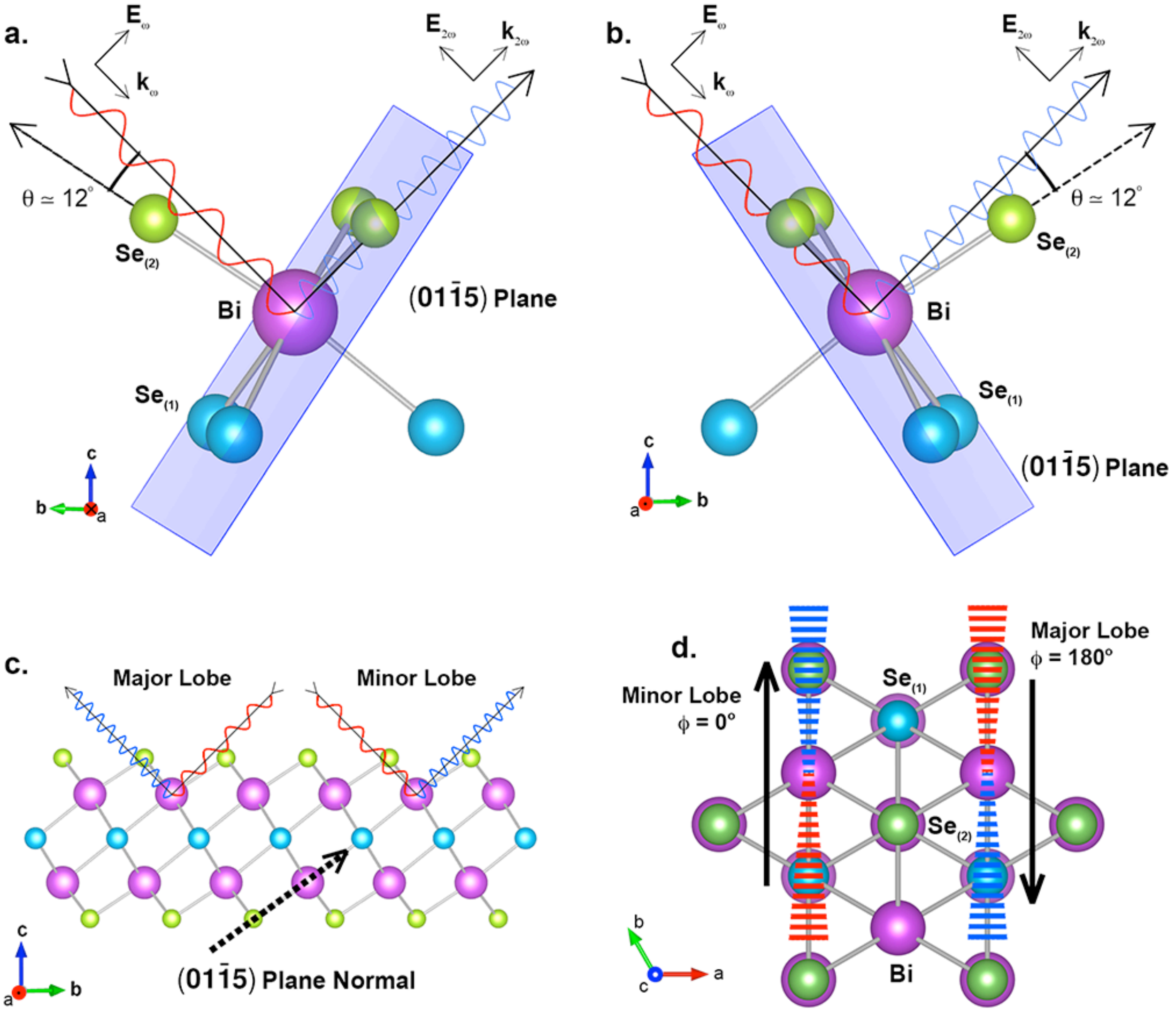

FIG. 11. Schematic of SHG at the surface of $Bi_2Se_3$ samples. In a freshly cleaved sample, each QL surface contains a Bi-$Se_{(2)}$ bond that points in the (0 1 $\bar{1}$ 5) plane normal direction. When the incoming ω light's propagation vector is nearly aligned with the (0 $\bar{1}$ 1 $\bar{5}$) plane normal direction (a), a minor SHG lobe is generated (0°, 120°, 240° in Fig. 10). As the sample is rotated around the *c* axis, the outgoing 2ω light's propagation vector becomes nearly parallel with the (0 1 $\bar{1}$ 5) plane normal direction (b), and a major SHG is generated (60°, 180°, 300° in Fig 10). These SHG orientations are shown in alternate views in (c) (side view along the *a* axis) and (d) (top-down view along the *c* axis).

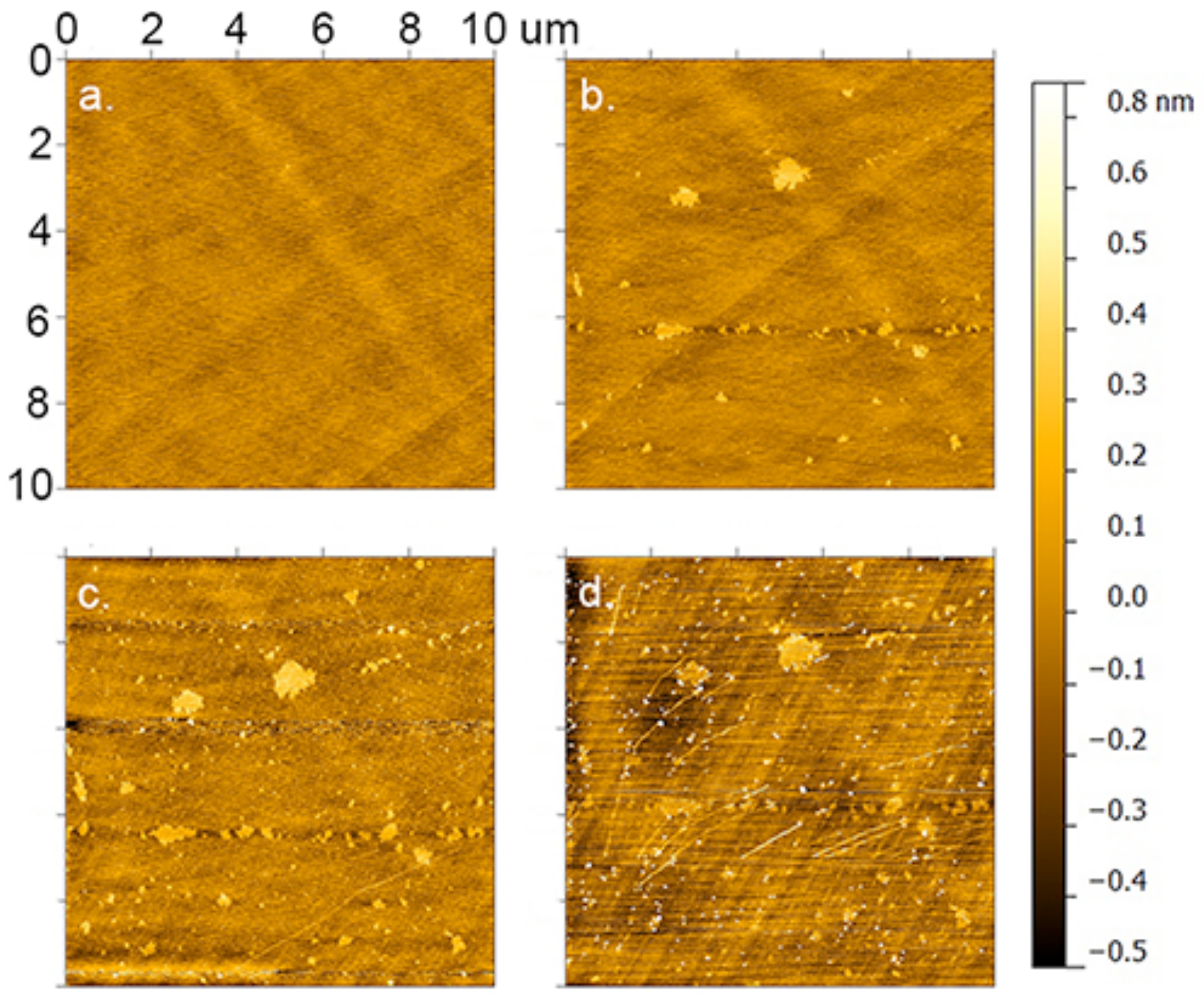

FIG. 12. Time-resolved AFM images of a clean surface (a), followed by patch growth (b-c), and then dirt adhesion (c-d). These patches, as in other measurements, have a uniform thickness of 3.6 ± 0.2 Å. The bright spots in (d) have thicknesses roughly equivalent to their diameters (>100 nm). Whereas the patches are quite large and flat, the dirt is roughly spherical, and contributes significantly to image aberrations.

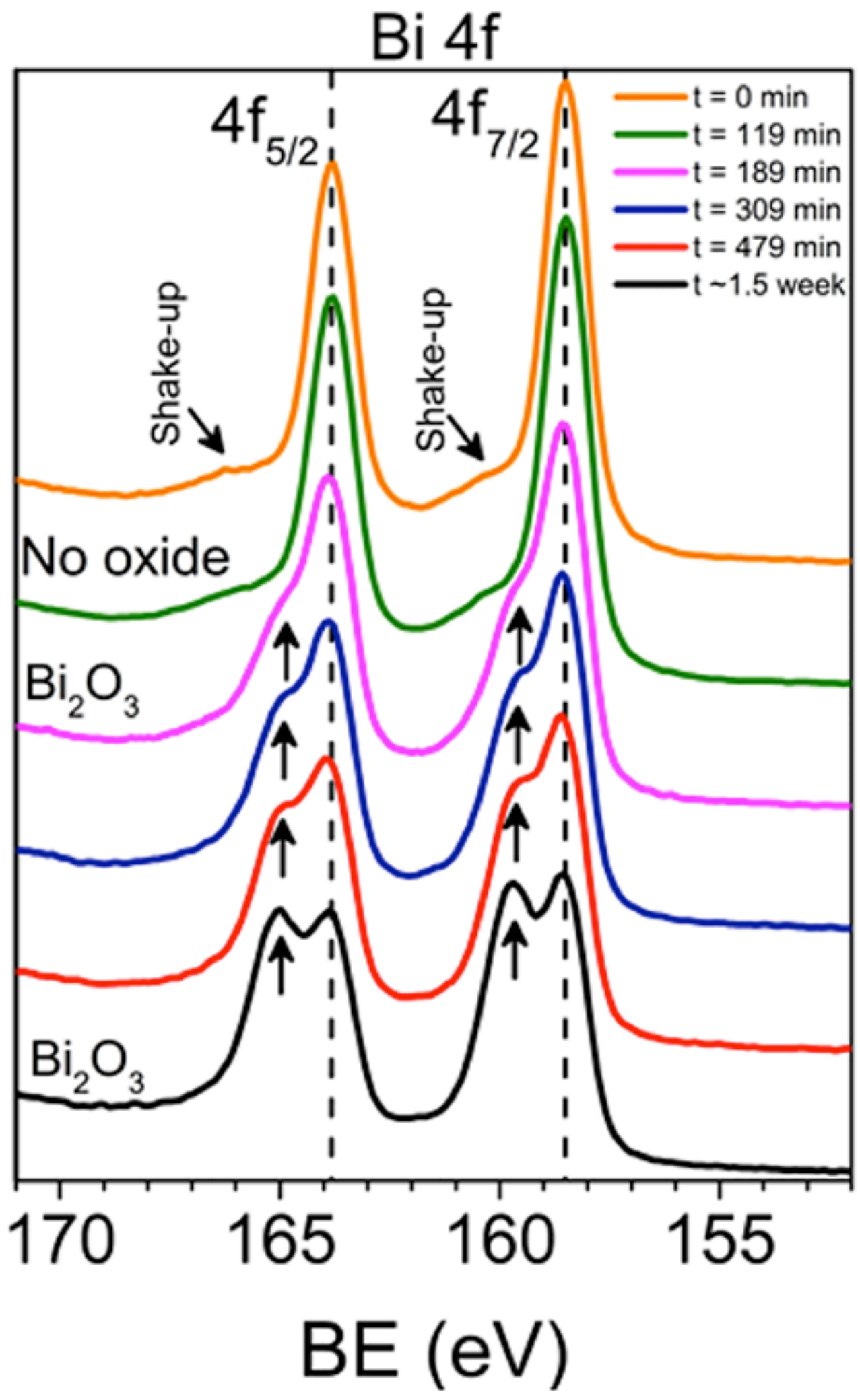

FIG. 13. XPS scans of the Bi 4f region taken over various times. Before t ~ 119 minutes, no oxidation is observed in the XPS data (see Fig. 5). As such, the presence of the shoulders to the left of the main Bi 4f peaks before t ~ 119 minutes are attributed to shake-up events[23,24] and overlapping peaks originating from Se[45]. For samples exposed to air for longer times, the Bi 4f oxide peaks also show up in the same region as that of the shoulders and grow in intensity over time.

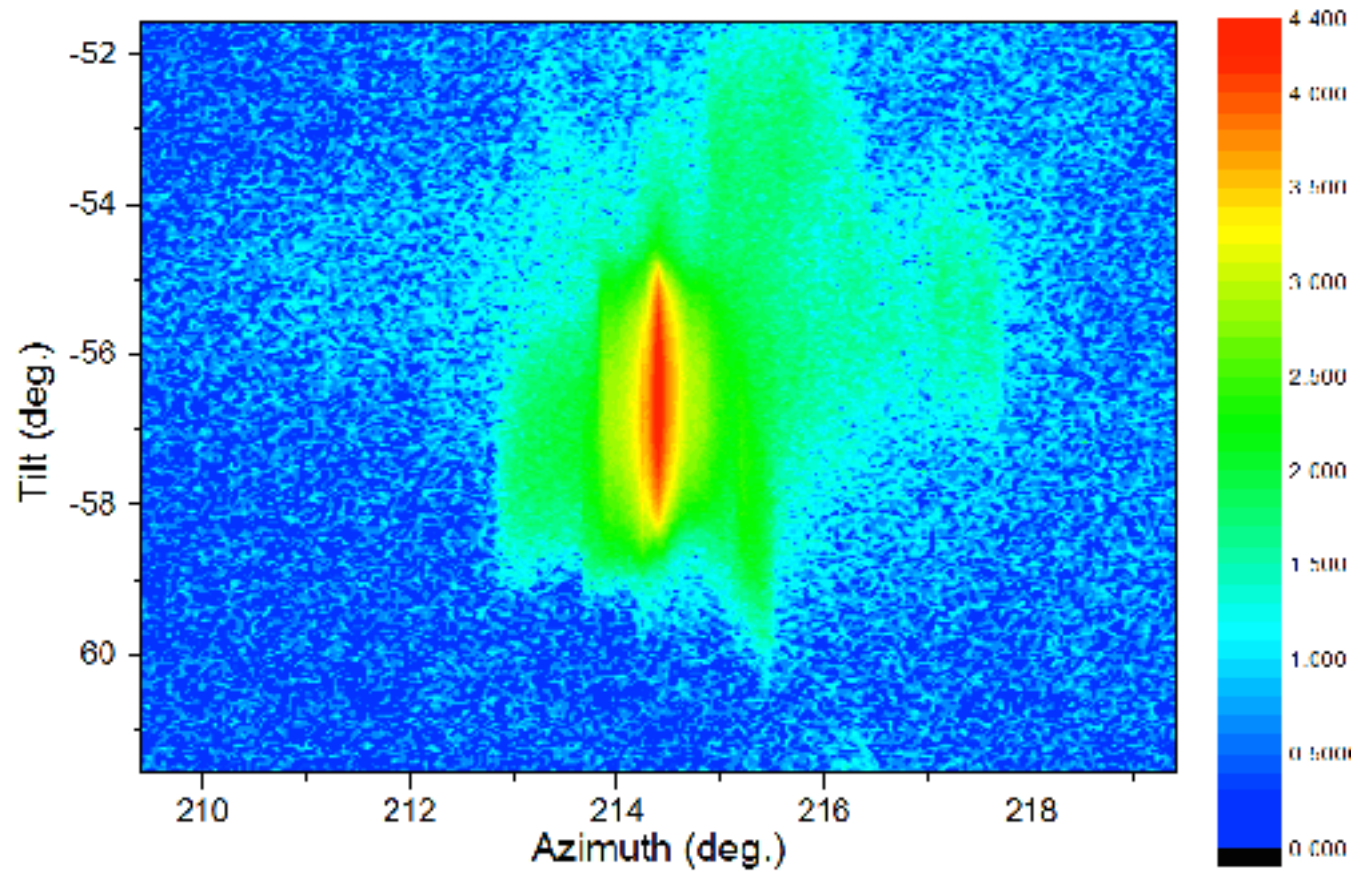

FIG. 14. Orientational map of a $Bi_2Se_3$ ($0\ 1\ \overline{1}\ 5$) peak. The contour axis represents the log of the detected x-ray intensity. This is one of the three-fold symmetric peaks (the two others having the same tilt, but azimuthal angles of ~ 94° and ~ -16°) that allowed SHG orientation to be determined conclusively. Additionally, the appearance of only three peaks indicates that our samples (~ 1 $cm^2$ area) were composed of single grains.